\documentclass{article}
\usepackage{amsmath}
\usepackage{graphicx}
\def\p{\partial}

\def\g{\gamma}

\def\de{\delta}
\def\D{\Delta}
\def\De{\Delta}
\def\ov{\overline}
\def\ld{\lambda}
\def\Ld{\Lambda}

\def\ep{\epsilon}

\def\e{\eta}
\def\et{\eta}

\def\om{\omega}
\def\Om{\Omega}

\def\b{\beta}

\def\a{\alpha}

\def\pdellx'{\frac{\partial}{\partial x'}}
\def\pdellw'{\frac{\partial}{\partial w'}}
\newcommand{\be}{\begin{equation}}
\newcommand{\ee}{\end{equation}}
\def\bed{\begin{displaymath}}
\def\eed{\end{displaymath}}
\def\bea{\begin{eqnarray}}
\def\eea{\end{eqncrray}}
\def\[{$$}
\def\]{$$}
\begin{document}
\title{A CONFINING QUARK MODEL AND \\ NEW GAUGE SYMMETRY}
\author{Jong-Ping Hsu\footnote{jhsu@umassd.edu}\\
Department of Physics, \\
University of Massachusetts Dartmouth, \\
North
Dartmouth, MA 02747-2300, USA}

\date{\today}
\maketitle
{\small  We discuss a confining model for quark-antiquark system with a new color $SU_3$ gauge symmetry.   New gauge transformations involve non-integrable phase factors and lead to the fourth-order gauge field equations and a linear potential.  The massless gauge bosons have non-definite energies, which are not observable because they are permanently confined in quark systems by the linear potential.    We use the empirical potentials of charmonium to determine the coupling strength of the color charge $g_s$ and find $g_s^2/(4\pi) \approx 0.2.$  We discuss Feynman-Dyson rules for the confining quark model, which involve propagators with poles of order 2 associated with new gauge fields.  The confining quark model may be renormalizable by power counting and compatible with perturbation theory.  }

\bigskip
%Keywords: quark confinement; new gauge symmetry; gauge transformation with non-integrable phase factor; fourth-order equation; linear %potential.

%PACS numbers:  11.30.-j,  11.15.-q. 

\bigskip
\section{Introduction }

In order to understand confining potentials for quarks, we generalize the usual $SU_3$ gauge transformations to new transformations involving a non-integrable phase factor.  This generalization leads to a new fourth-order gauge-invariant field equations.  The non-homogeneous field equation leads to a particular solution $\propto r$ and the homogeneous equation has a solution $\propto 1/r$.  Their coefficients are determined by physical boundary conditions.

We demonstrate that the conservation of the quark current can also be understood in terms of the new $(SU_3)_{color}$  gauge symmetry.   We replace the usual Lorentz scalar gauge function $\Ld^a(x)$ by a vector function $\om^a_{\mu}(x)$, which leads naturally to a non-integrable phase factor involving the vector gauge function $\om^a_{\mu}(x)$.  In the special case,  $\om^a_\mu(x) = \p_{\mu}\Ld^a(x)$,  the non-integrable phase factor reduces to the usual phase factor.  In the literature, Yang and Wu stressed that electromagnetism is the gauge invariant manifestation of the non-integrable phase factor $exp(ie\oint  A_{\mu} dx^{\mu})$, which provides an intrinsic and complete description of electromagnetism.\cite{1,2}   We used vector gauge functions and non-integrable phase factor to investigate forms of gauge fields, wrapping numbers and quantization conditions in gauge field theories.\cite{3}  In this paper, we further generalize the form of vector gauge functions and employ them to explore a confining  quark model, fourth-order gauge field equations\cite{4}  and confining potentials.   The massless gauge bosons associated with fourth-order equations have non-definite energies,\cite{5}  which turn out to be not observable because, in the present model, these gauge bosons are permanently confined in quark systems by the linear potential. 

The excited states of the charmonium provide empirical potentials between charmed quark-antiquark.\cite{6}  The potentials have a linear potential in $r$ and a Coulomb-type potential $\propto r^{-1}$.  It is interesting to note that the coupling strength associated with the Coulomb-type potential cannot be purely electromagnetic.  The presence of both the linear and the Coulomb-type potentials are just right for the general solution of the fourth-order gauge field equation with new $(SU_3)_{color}$ gauge symmetry. 
  
 \section{New Color $SU_3$ Gauge Transformations }
 
Suppose  gauge fields $H^a_\mu (x)$ and quark fields $q(x)$ are associated with $(SU_3)_{color}$.  
For quarks, the generalized gauge transformations involving  vector gauge functions $\om^a_{\ld}(x)$ are defined by
\be
q'(x) = \Om(x) q(x), \ \ \ \ \   \ov{q}'(x) =\ov{q}(x)\Om^{-1}(x),    
\ee
%%%%%2%%%1
\be
 \Om(x)=exp\left[- i g_s \int^{x}  \om_{\ld}(x') dx'^\ld \right],
\ee
%%7%%%%%%%3%%%%2
\be
 \om_\ld(x)=\om^a_{\ld}(x)L^a,   \ \ \ \  [L^a , L^b]= if^{abc} L^c, \ \ \ \   L^a=\frac{\ld^a}{2}.
\ee
%%%%%%%%3
where $c=\hbar=1$ and  $\ld^a$, a=1,2,...8, are Gell-Mann matrices.\cite{7}  The path in (2)  could be arbitrary, as long as it ends at the point $x\equiv x^\nu$.  
 
  The new gauge transformations for  $H^a_\mu(x)$ are defined by  
 \be
H'_\mu(x) = \Om(x)H_\mu(x)\Om^{-1}(x)- \frac{i}{g_s}\Om(x)\p_\mu \Om^{-1}(x),
 \ee
%%%%%%4
where $H'_\ld(x)=H'^a_{\ld}(x)L^a.$
The  $SU_3$ gauge covariant derivatives $\De_\mu$ and gauge curvatures $H_{\mu\nu}(x) $ are defined as usual,
\be
\De_\mu=\p_\mu + ig_s H^a_\mu (x) L^a,  
\ee
%%%4%%%%5
\be
[\De_{\mu} , \D_\nu]=ig_s H_{\mu\nu}(x),   \ \ \ \ \ \ \   H_{\mu\nu}(x) = H^a_{\mu\nu} (x) L^a,
\ee
%%%%%%15%%%%13%%%%5%%%6
where
%\be
%H^a_{\mu\nu}(x) = \p_\mu H^a_\nu(x) - \p_\nu H^a_\mu(x) - f f^{abc} H^b_\mu(x)H^c_\nu(x),
%\ee
%%%%%28%%532%%%%%59%%557%%%%%40%%%38%%%16%%%14%%%11%%%%6
\be
H_{\mu\nu}(x) = \p_\mu H_\nu(x) - \p_\nu H_\mu(x) + i g_s [H_\mu(x), H_\nu(x)].
\ee
%%6%%7
We have the following new gauge transformations for  $H_{\mu\nu}(x)$ and $\p^\mu H_{\mu\nu}(x)$ :
\be
H'_{\mu\nu}(x) = H_{\mu\nu}(x)+\p_\mu \om_\nu(x) - \p_\nu \om_\mu(x)  -ig_s \left[\int^{x}\om_{\ld}(x') dx'^{\ld},  H_{\mu\nu}(x)\right],  
\ee
%8%%%%28%%27%%526%%%%%%%%13%%%%9%%%8%%%%16%%%13%%%8
for infinitesimal $\om_\mu(x)$.  However, $\p^\mu H_{\mu\nu}(x)$ transforms properly under (4),
\be
\p^\mu H'_{\mu\nu}(x)= \p^\mu H_{\mu\nu}(x) -ig_s \left[\int^{x}\om_{\ld}(x') dx'^{\ld}, \p^\mu H_{\mu\nu}(x)\right], 
\ee
%%%%9%%29%%28%%%%27%%%%%%%%%%14%%%%10%%%9%%%18%%%%15%%%9
provided the gauge functions $\om_\mu (x)$ satisfy the following constraints
\be
\p^\mu \{\p_\mu \om_\nu (x) - \p_\nu \om_\mu (x)\} -ig_s [\om^\mu(x) , H_{\mu\nu}(x)] = 0.
\ee
%%%%%%%%%%%%19%%%%%17%%%10
The constraints in (10) are necessary for the gauge invariance of the Lagrangians with  $(SU_3)_{color}$ group.   We stress that the Lagrangian with new $SU_3$ gauge invariance can be constructed only for the general gauge function, $\om_{\mu}(x) \ne \p_{\mu} \om(x)$.   The restriction for $\om_\mu(x)$ in (10) is similar to that for gauge functions of Lie groups in the usual non-Abelian gauge theories.\cite{8,9}  We also verify that $ \ov{q}\g^\mu\De_\mu q$ is invariant  under the transformations  (1) and (2), where we have used the relation $
\p_\mu  \Om_{\om}(x) =  - i g_s  \om^a_\mu (x) L^a \Om_{\om}(x)$.

\section{A New Invariant Lagrangian and Confining Potentials}
In the confining model for quark-antiquark system, the  Lagrangian with new $[SU_3]_{color}$ gauge symmetry takes the form
\be
L_{qH}=\frac{1}{2}L_s^2 \p_{\mu} H^{\mu\ld a}\p^{\nu} H^a_{\nu\ld} + \ov{q}(x) [i\g^\mu(\p_\mu+ig_s H^a_{\mu} L^a)- M]q(x),
\ee
%%41%%%%%%%40%%%%18%%%12%%%11
where $M$ is the mass matrix of quarks.\cite{7}  The gauge field equation for $H^a_\mu(x)$ can be derived from the Lagrangian (11),
\be
\p^2 \p^\mu H^a_{\mu\nu} -\frac{g_s}{L_s^2}\ov{q} \g_\nu L^a q = 0.
\ee
%%12%%34%%%%38%%%19%%%%24%%13%%%%12
One can also derive quark equations from the gauge invariant Lagrangian (11) and obtain the continuity equation $
\p^\nu(\ov{q} \g_\nu L^a  q )=0$
associated with the color  $SU_3$ group.  The new gauge symmetry and the resultant gauge fields satisfying the fourth-order equations may be called respectively `taiji gauge symmetry'\cite{4} and `taiji gauge fields,' in order to distinguish them from  usual gauge symmetries and gauge fields.  The massless gauge bosons associated with $H^a_\mu(x)$ may be called 'confions.' 

Since there is a good empirical potential obtained by fitting the spectrum of charmonium, let us consider the static solution of (12) produced by charmed quarks, $q(x)=c(x)$.  We are interested in the static equation for the time-component  $H^a_0({\bf r})$ in (12), 
\be
L^2_s \nabla^2 \nabla^2 H^a_0 ({\bf r})= g_s \ov{c}\g_0 L^a c, \ \ \ \ \   a=1,2,3,.....8.
\ee
%%%%%%23%%%%28%%%27%%%13
To be specific, let us first consider $a=8$. For a spherical symmetric and static solution, we replace the time component of the source term in (13) by a point source at the origin.
\be
\nabla^2 \nabla^2 H^8_0 ({\bf r})= \frac{g_s}{L^2_s} \ov{c}\g_0 L^8 c = \frac{g_s}{2\sqrt{3}L^2_s}[\de^3({\bf r})_{11} + \de^3({\bf r})_{22} -2 \de^3({\bf r})_{33}].
\ee
%14
To avoid confusion, we use  $\de^3({\bf r})_{kk}$ to denote the usual delta function $\de^3({\bf r})$ associated with the source $ \ov{c}_k\g_0 c_k $ with k=(1,2,3) denoting the color (red, blue, green) for three different color quarks.\cite{7}  In other words, the source $ \ov{c}_1\g_0 c_1 $ is replaced by a point red charge at the origin,  denoted by $\de^3({\bf r})_{11}$, etc.  Each of the three source terms in (14) corresponds to a specific Feynman diagram involving a color charge, which can be expressed in units of  $g_s$.  The potential due to the exchange of virtual confion corresponding to $H^8_0$ between $\ov{c}_1$ and $c_1$ satisfies the equation,
\be
\nabla^2 \nabla^2 H^8_0 ({\bf r})= \frac{Q_r}{L_s^2} \de^3({\bf r})_{11}, \ \ \ \ \ \   Q_r =\frac{g_s}{2\sqrt{3}}.
\ee
%15
This equation has a particular solution
\be
H^8_0({\bf r}) = {-Q_r r}/{8\pi L^2_s}, 
\ee
%16
where we have used the Fourier transform of generalized functions.\cite{10,4}  For general solutions, we also need a solution $H'^8_0(r)$ for the homogeneous equation, i.e., (15) without the source term.  We observe that this solution is
$$
 H'^8_0({\bf r})=b/r, 
 $$
 which satisfies $\nabla^2 H'^8_0 ({\bf r})=0$ for $r \ne 0$.  To determine the unknown constant $b$, we require that the derivative of the solution of the homogeneous equation satisfies the physical boundary condition, $\nabla^2 H'^8_0 (0)=- Q_r \de^3(0)$, at the origin.  This boundary condition implies that the interaction of taiji gauge fields is characterized by one single red charge $Q_r$, i.e., $b/r= Q_r/(4\pi r)$.  Thus the general static solution leads to a `dual potential' with the  potential energy $ - Q_r (H^8_0+H'^8_0)$ for the $\ov{c}_1 c_1$ system.  We should also consider the attractive electromagnetic potential energy between charmed quark and anti-quark with the charges $Q_{e1}=2e/3$ and $Q_{e2}=-2e/3$, i.e., $V_{em}= {Q_{e1} Q_{e2}}/(4\pi r)$.  Thus, the potential energy for the $\ov{c}_1 c_1$ system is
\be
V_{11}=\frac{Q_r^2 r}{8 \pi L_s^2} - \frac{Q_r^2}{4\pi r}, \ \ \ \  V_{em}= - \frac{ e^2}{9 \pi r}, \ \ \     Q_r=\frac{g_s}{2\sqrt{3}}.
\ee
%%%%%%%%%43%%%%26%%%31%30%%%%%15%%514%%%17
 Similarly, the potential energies for the  $\ov{c}_2 c_2$ and the $\ov{c}_3 c_3$ systems are
\be
V_{22}=\frac{Q_b^2 r}{8 \pi L_s^2} - \frac{Q_b^2}{4\pi r}, \ \ \ \ \ \ \    Q_b=\frac{g_s}{2\sqrt{3}},
\ee
%18
\be
V_{33}=\frac{Q_g^2 r}{8 \pi L_s^2} - \frac{Q_g^2}{4\pi r}, \ \ \ \ \ \ \    Q_g=\frac{-g_s}{\sqrt{3}},
\ee
%%%%%19
Thus, the sub-total potential energy for the charmonium mediated by the $H^8_\mu$ is given by the sum of (17), (18) and (19),
\be
V_{8st}=\frac{g_s^2}{2}\left (\frac{ r}{8 \pi L_s^2} - \frac{1}{4\pi r}\right).
\ee
%%%%%%20
One can also obtain the sub-total potential energy for the charmonium mediated by the $H^3_\mu$ field,
\be
V_{3st}=\frac{g_s^2}{2}\left (\frac{ r}{8 \pi L_s^2} - \frac{1}{4\pi r}\right),
\ee  
%%%%%%%%21
where the static gauge field equation for $H^3_0$ is
\be
\nabla^2 \nabla^2 H^3_0 ({\bf r})= \frac{g_s}{2L^2_s} \ov{c}\g_0 L^3 c = \frac{g_s}{2L^2_s}[\de^3({\bf r})_{11} - \de^3({\bf r})_{22}].
\ee
%%%%%22
Next, let us consider the exchange of color confions corresponding to $X_\mu =(H^1_\mu + i H^2_\mu)/\sqrt{2}$,  $Y_\mu =(H^4_\mu + i H^5_\mu)/\sqrt{2}$,   $Z_\mu =(H^6_\mu + i H^7_\mu)/\sqrt{2}$, and their complex conjugates.\cite{7}  Eq. (13) gives
\be
\nabla^2 \nabla^2 X_0 ({\bf r})= \frac{g_s}{\sqrt{2}L^2_s} \ov{c}_1 \g_0 c_2 = \frac{g_s}{\sqrt{2}L_s^2} \de^3({\bf r})_{12},
\ee
%%%%%%%23
and its complex conjugate,
\be
\nabla^2 \nabla^2 X^*_0 ({\bf r})= \frac{g_s}{\sqrt{2}L^2_s} \ov{c}_2 \g_0 c_1 = \frac{g_s}{\sqrt{2}L_s^2} \de^3({\bf r})_{21}.
\ee
%%%%%%%%%24
Following the previous steps of calculations from (15) to (20), one can also obtain the sub-total potential energy for the charmonium mediated by the $X_\mu$ and $X^*_\mu$,
\be
V_{Xst}= g_s^2\left (\frac{ r}{8 \pi L_s^2} - \frac{1}{4\pi r}\right).
\ee
%%%%%%%25
 Similarly, we also have
\be
V_{Yst}= V_{Zst}= g_s^2\left (\frac{ r}{8 \pi L_s^2} - \frac{1}{4\pi r}\right),
\ee
%%%%%%%%%26
which was contributed by the fields $Y_\mu. Y^*_\mu, Z_\mu$ and $Z^*_\mu$.
The potential energies for charmonium can be separated into two parts: (a) the sum of  (20) and (21), including $V_{em}$, and (b) the sum of (25) and (26).  We have
\be
 V_{tot}= g_s^2 \left (\frac{ r}{8 \pi L_s^2} - \frac{1}{4\pi r}\right) - \frac{4 e^2}{9}\left(\frac{1}{4\pi r}\right),
 \ee
 %%%%%%%%%%%27
$$
V_{XYZ} = V_{Xst}+V_{Yst}+V_{Zst}= 3 g_s^2 \left (\frac{ r}{8 \pi L_s^2} - \frac{1}{4\pi r}\right),
$$
where only $V_{tot}$ is similar to the usual potential, in which quarks do not change color due to the emission and absorption of a virtual confion.
To calculate the constants $L_s$ and color charge $g_s$ in the Lagrangian (11), we use the empirical potential energy $V(r)$ of the Cornell group obtained by fitting the spectrum of charmonium\cite{6},
\be
V(r) = -\frac{\a_c}{r}\left[1-\left(\frac{r}{a}\right)^2 \right],  
\ee
%%%28%%%%33%%%32%%%%15%%%%28
$$
 \ \ \  \a_c=0.2, \ \ \ \ \   a=0.2 fm, \ \ \ \   m_c = 1.6 GeV.
 $$
  In the approach of Cornell group for charmonium spectrum, heavy charmed quarks are considered as non-relativistic particles.
It is natural to identify the empirical potential $V(r)$ in (28) with the total confining potential energy (27), including the electromagnetic potential energy.
Thus, we obtain the coupling strength $g_s^2/(4\pi)$ and the confining scale length $L_s$ to be 
\be
\frac{g_s^2}{4\pi} \approx 0.2, \ \ \ \ \ \   L_s  \approx 0.14 fm, \ \ \ \ \  m_c \approx1.6 GeV.
\ee
%%%%%%%%17%%%%30%%%%29
Since  $m_c=1.6 GeV$ differs from the present experimental values,\cite{11} $1.18 GeV \le m_c \le 1.34 GeV$, with an uncertainty of roughly $30 \%$, other values in (30) may have similar uncertainties.  Nevertheless, it is interesting that the coupling strength ${g_s^2}/(4\pi) $ in the confining model is still much smaller than 1.  This result suggests that the confining quark model may be consistent with perturbation theory.

\section{New Feynman-Dyson rules for the Confining Quark Model }

Since the generalized gauge transformations involve non-integrable phase factor (2) and  the vector gauge functions, it appears necessary to modify Faddeev-Popov method\cite{12} in order to the quantize taiji gauge fields and to remove unphysical amplitudes due to the interaction of the unphysical longitudinal and time-like components\cite{13} of the massless  confion.  

We assume that the Faddeev-Popov method can be suitably modified and applied to the more complicated taiji $SU_3$ symmetry involving the fourth-order gauge field equations.    To quantize the confining model and to obtain new Feynman-Dyson rules for Feynman diagrams in the confining model, the total Lagrangian $L_{tot}$ must including the gauge fixing term  $L_{gf}$.   We assume $L_{tot}$ to be 
\be
L_{tot} = L_{qH} + L_{gf}, \ \
\ee
%%%%%%%30
$$
 L_{gf}= \frac{L_s^2}{2\a}(\p_\ld \p_\mu H^{\mu a})(\p^\ld \p^\nu H^a_{\nu}),
$$
%%%
which corresponds to the choice of a new gauge condition
\be
\p_\ld \p_\mu H^{\mu a}=y^a_\ld(x),
\ee
%%%%%19%%%%%%%31
where $y^a_\ld(x)$ is a suitable function independent of the fields and the vector gauge functions.  It can be verified that the gauge condition (31) cannot be imposed for all times because  the unphysical components of the taiji gauge field $\p_\mu H^{\mu a}$ do not satisfy the free equation, in contrast to that in quantum electrodynamics.  The vacuum-to-vacuum amplitude of the model is\cite{12,13} 
$$
 W(y^a_\ld)= \int d[H,q,\ov{q}] exp\left(i \int L_{qH} d^4 x\right) 
 $$
 \be
\times det Q  \prod_{a,x,\ld} \de(\p_\ld \p_\mu H^{\mu a}-y^a_\ld),
 \ee
 %%%20%%%32
where the functional determinant detQ can be calculated by considering the infinitesimal  taiji gauge transformation of (31), i.e., $\p_\ld \p_\mu H'^{\mu a}-y^a_\ld=0$\cite{12,14}.  The generating functional $W(J)$ in the gauge specified by (31) is given by
\be
 W(J)= \int d[H,D,\ov{D},q,\ov{q}] exp\left(i \int (L_{eff}+J^a_\mu H^{a \mu} + \ov{q}\et + \ov{\et}q) d^4 x\right),
 \ee 
 %%%21%%%33
 \be
L_{eff}= L_{tot} + L_D, \ \ \   L_D=L_s^2[\p^2 \ov{D}^a \p^2 D^a - g_s f^{abc} (\p^2 \ov{D}^a)\p_{\mu} (H^{\mu b} D^c)].
\ee
%%22%%%34
 The ghost Lagrangian $L_D$ contains all interaction terms corresponding to those of  the longitudinal and time-like confion.  In the derivation of (21) and (22) based on $\p_\ld \p_\mu H'^{\mu a}-y^a_\ld(x)=0$, the infinitesimal vector gauge function $\om^a_{\mu}$ is assumed to correspond to the derivative of ghost fields, $\p_\mu D^a$, so that  the ghost fields $D^a$ and $\ov{D}^a$ satisfy fourth-order equations.  As usual, they are quantized as scalar fermions.   In this way, they should  produce amplitudes to cancel completely the extra unwanted amplitudes due to the interactions of unphysical longitudinal and time-like confions.  Thus, gauge invariance and unitarity of the S-matrix in the confining model can be restored.  Since this modified Faddeev-Popov method is not completely the same as those in the usual gauge theories, it is desirable that the unitarity and gauge invariance of the S-matrix  be independently verified in the future by explicit calculations\cite{15} of the imaginary parts of amplitudes of physical processes based on $L_{eff}$ in (34).

The complete Feynman-Dyson rules in the confining model for Feynman diagrams  can be derived from the effective Lagrangian $L_{eff}$ in (34):  The confion and ghost propagators are respectively given by
 \be
 C^{ab}_{\mu\nu}(k)=\frac {-i\de^{ab}}{L^2_s (k^2+i\ep)^2}\left[\e_{\mu\nu} -(1-\a)\frac{k_\mu k_\nu}{k^2+i\ep}\right],
 \ee
 %%%46%%%%29%%%%534%%%%%19%%%23%%%35
 $$
  G^{ab}(k)=\frac {-i\de^{ab}}{L^2_s (k^2+i\ep)^2},
  $$
 which involve a pole of order 2 rather than a simple pole due to the higher order derivatives in (34), in contrast to the photon propagator.  
 The 3-confion vertex $[H^a_\a(k_1)H^b_{\b}(k_2)H^c_{\g}(k_3)]$ is given by
$$
g_s L^2_s f^{abc}[(k_1)^2( k_{2\b}\e_{\a\g}- k_{3\g}\e_{\a\b} + k_{3\b}\e_{\a\g} - k_{2\g}\e_{\a\b})
$$
$$
+ (k_2)^2( k_{3\g}\e_{\a\b}- k_{1\a}\e_{\g\b} + k_{1\g}\e_{\a\b} - k_{3\a}\e_{\g\b})
$$
\be
+ (k_3)^2( k_{1\a}\e_{\b\g}- k_{2\b}\e_{\a\g} + k_{2\a}\e_{\b\g} - k_{1\b}\e_{\a\g})],
 \ee
 %%%%%22%%24%%%36
where $(k_n)^2=k_{n\mu}k_n^\mu, \  n=1,2,3$.  The expression for the 4-confion vertex $[H^{a}_\a(k_1)H^{b}_\b(k_2) 
H^{c}_\g(k_3)H^{d}_\de(k_4)]$ is more complicated.  For the discussion of the degree of divergence associated with (34), it suffices to give some typical terms,
$$
 ({-i}/{2})L^2_s g^2_s[f^{nac}f^{nbd}(k_{1\a}k_{2\b}\e_{\g\de}- k_{1\a}k_{4\de}\e_{\g\b})
 $$
 $$
+ f^{nad}f^{nbc}(k_{1\a}k_{2\b}\e_{\g\de}- k_{1\a}k_{3\g}\e_{\b\de})
 $$
 \be
+ f^{nab}f^{ncd}(k_{1\a}k_{3\g}\e_{\b\de}- k_{1\a}k_{4\de}\e_{\g\b})+......].
 \ee
 %%%%%23%%25%%%37
The confion-ghost vertex involves $k^3$ terms.  The momenta are in-coming to the vertices.  The  rules for quark propagator and the confion-quark vertex  
[$\ov{q}(k) q(p) H^a_{\mu}(q)$], are given by $i/(\g^\mu p_\mu - m +i\ep)$ and $ -i g_s \g_{\mu}L^a$ respectively. 
Other rules\cite{16} such as a factor -1 for each fermion loop, a symmetry factor (1/2!) for the confion self-energy and for tadpole diagrams are the same as the corresponding diagrams in quantum chromodynamics (QCD).\cite{17}

From (34)-(37), we have seen that the confion and ghost propagators are more convergent at high energies than the corresponding propagators in usual gauge theory by a factor $k^{-2}$, while the 3-confion, 4-confion and confion-ghost vertices are all more divergent than the corresponding QCD vertices by a factor $k^2$.  To be specific, one can check the one-loop diagrams with non-negative superficial degree of divergence such as (A) confion, ghost and quark self-energy parts and (B) 3-confion, 4-confion, confion-ghost and confion-quark vertices.\cite{17}  All these divergences are no worse than the corresponding diagrams in QCD.
Thus, the confining model could be renormalizable by naive power counting and with the dimensional regularization.  

There is also a linear potential between the quark and the massless confion, which is due to the exchange of a virtual confion between them.  Because of the k-square terms in 3-confion vertex (36), one expects that  strong dual potential $V$ in (27)  arises from the exchange of massless confions: $V=(2\pi)^{-3}\int M_{fi} exp(i{\bf k\cdot r}) d^3k$, where the potential is just the 3-dimensional Fourier transform of the lowest-order $M$-matrix element.\cite{18}  In comparison with other models for confining quarks such as lattice QCD and potential NRQCD,\cite{19,20,21} which involve elaborated computations and approximations, the present confining model has the advantage of simplicity in the derivation of confining potential and of close connection to a new gauge symmetry.
 
 The small value for the strength of color charge in (29) suggests that perturbation calculations could be useful in this confining model with taiji gauge symmetry.   
The empirical potentials indicates the necessity of both a linear potential in $r$ and a non-electromagnetic Coulomb-type potential $\propto r^{-1}$.  These properties are just right for the dual potential solution (27)  of the fourth-order gauge field equation with taiji $(SU_3)_{color}$ gauge symmetry. 
   The fourth-order gauge field equations with taiji gauge symmetry play a crucial role in  qualitative understanding of the confinement of quarks and confions in  quark-antiquark system.   We note that the construction of a model with an attractive force for all quarks and antiquarks appears to be beyond the conceptual framework of the conventional  gauge symmetry with internal groups.  It seems likely that the taiji gauge symmetry will also play a role for understanding baryon (or three-quark) systems. Since the forces between three quarks are generated by color isotopic charge and color hypercharge,\cite{7} they involve repulsive and attractive forces.  It is more difficult to understand  three-quark systems such as protons and neutrons, in contrast to the quark-antiquark systems.  Nevertheless, one may speculate that the strong force between baryons may resemble the chemical bonds between atoms.\footnote{The author would like to thank L. Hsu for interesting discussions.}
      
   In conclusion, it appears that quark confinement is the gauge invariant manifestation of a non-integrable phase factor, which leads to quark dynamics with taiji gauge symmetry and confining dual potentials.

\bigskip

{\bf ACKNOWLEDGMENTS}

The author would like to thank his colleagues for discussions.  The work was support in part by the Jing Shin Research Fund, UMass Dartmouth Foundation.

\newpage  
\bibliographystyle{unsrt}

%{99}

\end{document}